\documentclass[a4paper]{article}
\usepackage[utf8]{inputenc}

\usepackage{graphicx}
\usepackage[utf8]{inputenc}
\usepackage[numbers]{natbib}
\bibpunct{[}{]}{,}{n}{}{;}
\usepackage{amsmath}
\usepackage{amssymb}
\usepackage{geometry}
\usepackage{authblk}
\usepackage{tabu}
\begin{document}

\title{Artificial Intelligence for Interstellar Travel}


\author[1]{Andreas M. Hein}
\affil[1]{Initiative for Interstellar Studies, Bone Mill, New Street, Charfield, GL12 8ES, United Kingdom}

\author[2]{Stephen Baxter}
\affil[2]{c/o Christopher Schelling, Selectric Artists, 9 Union Square 123, Southbury, CT 06488, USA.}

\maketitle

\begin{abstract}
The large distances involved in interstellar travel require a high degree of spacecraft autonomy, realized by artificial intelligence. The breadth of tasks artificial intelligence could perform on such spacecraft involves maintenance, data collection, designing and constructing an infrastructure using in-situ resources. Despite its importance, existing publications on artificial intelligence and interstellar travel are limited to cursory descriptions where little detail is given about the nature of the artificial intelligence. This article explores the role of artificial intelligence for interstellar travel by compiling use cases, exploring capabilities, and proposing typologies, system and mission architectures. Estimations for the required intelligence level for specific types of interstellar probes are given, along with potential system and mission architectures, covering those proposed in the literature but also presenting novel ones. Finally, a generic design for interstellar probes with an AI payload is proposed. Given current levels of increase in computational power, a spacecraft with a similar computational power as the human brain would have a mass from dozens to hundreds of tons in a 2050 – 2060 timeframe. Given that the advent of the first interstellar missions and artificial general intelligence are estimated to be by the mid-21st century, a more in-depth exploration of the relationship between the two should be attempted, focusing on neglected areas such as protecting the artificial intelligence payload from radiation in interstellar space and the role of artificial intelligence in self-replication. 

\textbf{Key words:} Interstellar travel, artificial intelligence, artificial general intelligence, space colonization

\end{abstract}

\section{Introduction}

Robotic deep space exploration and interstellar travel require high levels of autonomy, as human intervention is very limited with signals taking years to travel to the probe and back. Autonomy is required for exploring the target star system, developing an infrastructure using local resources, and even colonization \cite{Hein2014a, Hein2014d, ArXiv:1612.08733}. High levels of autonomy in spacecraft are associated with performing cognitive tasks such as image recognition, reasoning, decision-making etc. For example, current planetary rovers are able to autonomously identify scientifically interesting rock formations via feature recognition and take decisions to analyze them \cite{Chien2017,ThompsonD.R.Wettergreen2011, Castano2007, Ingrand2007, Woods2009}. A program that is able to perform such and other cognitive tasks is referred to as artificial intelligence (AI) in the following \cite{Hernandez-Orallo2017}. 

An overview of the current state of the art of artificial intelligence in space exploration has been provided in Chien et al. \cite{Chien2006} and Chien and Wagstaff \cite{Chien2017}. According to Chien and Wagstaff \cite{Chien2017}, the main goals of AI on space probes is to detect and characterize features of interest such as usual and static (snow, water, ice, etc.) or unusual and dynamic (volcanic activity, fires, floods, dust devils, active jets), autonomous collection of interesting samples, autonomous creation of environmental maps, on-board analysis of data is desirable for reducing the data that needs to be stored and transmitted, on-board scheduling where mission scheduling needs to be adapted to unexpected events. Satisfactory schedules are searched based on a timeline model of the spacecraft state and its resources using AI. Interesting areas of future development are collaborative spacecraft and rovers that use sensor webs fusing data from various sensors \cite{Bartsch2010,Chien2017,Saeedi2016}. 

Chien et al. \cite{Chien2006} and Hein et al. \cite{Hein2012b} further explore the role of AI in human space exploration. AI-based mission operations scheduling can help crews to interactively schedule their activities \cite{Chien2006}. Managing the spacecraft systems such as the power subsystem is also considered an area where AI can support humans, in particular in off-nominal situations. In such a case, AI can perform problem analysis, perform repair actions, and evaluate the impact on future operations \cite{Chien2006,Hein2012b}. Furthermore, operations on planetary surfaces is another area where AI can assist in developing operations plans. 

Looking into the far future, robotic probes with sophisticated artificial intelligence capabilities have been proposed self-replicating space probes and probes that are capable of communicating with extraterrestrials \cite{Armstrong2013,Baxter2013,Bracewell1960,Dick2003,Tipler1986,Tipler1994}. The scenario of an interstellar probe encountering an extraterrestrial intelligence has been explored by Baxter \cite{Baxter2013}. Bacewell \cite{Bracewell1960} proposed an intelligent interstellar probe, a so-called Bracewell probe, that is able to perform sophisticated communication with extraterrestrials and contains large amounts of knowledge of a civilization. Combining the Bracewell probe with a self-replicating capability has been explored by Freitas \cite{Freitas1980a}, O'Neill \cite{ONeill1981} and Jones \cite{Jones2017}. Such advanced probes would require levels of artificial intelligence that are similar to human intelligence or even superior in broad task categories \cite{Everitt2018}. An artificial intelligence that is able to perform a broad range of cognitive tasks at similar levels or better than humans is called artificial general intelligence (AGI). Estimates for the advent for AGI differ \cite{Everitt2018}, however, their median is somewhere in the middle of the 21st century. The estimated launch date for the first interstellar probe falls into a similar time frame. Given these estimates, it is plausible to assume that AGI and interstellar travel might materialize at similar points in time and implications of one on the other are worth to be considered. 

The more mundane use of AI for maintenance and housekeeping of interstellar probes and crewed interstellar spacecraft has been explored for the Project Daedalus study  \cite{Bond1978} and world ships \cite{Hein2012b}. Precursors for such technologies have been developed in the context of using augmented reality and intermediate simulations for space stations \cite{Kirchner2014,Kirchner2015,Kroemer2017,Kroemer,Kroemer2016,Wirkus2014}. 

Past publications have mostly dealt with the principle feasibility of interstellar probes with an AI without providing engineering details of such a probe. For example, Tipler assesses the principle feasibility of mind-uploading into an artificial substrate and how a fusion-propelled Daedalus-type interstellar probe could transport the AGI to other stars and gradually colonize the universe \cite{Bond1978,Tipler1986}. Ray Kurzweil in "The Singularity is Near" describes nano-probes with AI payloads that could even traverse small worm holes for colonizing the universe \cite{Kurzweil2005}. The most sophisticated analysis of AI probes is provided by Bradbury who introduces the concept of "Matrioshka Brain" where a large number of spacecraft, producing power for AIs, orbit a star \cite{Bradbury2001}. Bradbury imagines whole layers of orbital rings around a star harnessing its energy, similar to a Dyson Sphere \cite{Badescu1995}. Hein introduces several potential mission architectures based on AI interstellar probes with the main objective of paving the way for human interstellar colonization by creating space or surface colonies in advance to their arrival \cite{Hein2014a,Hein2014d}. Using AGI to grow and raise humans from individual human cells or embryos at another star and thereby avoiding the transport of grown-up humans has been proposed by Crowl et al. \cite{Crowl2012}. 

Regarding mid- and far-term prospects of AI in space applications, these can be categorized into building artifacts in space, communicating with extraterrestrials, and growing / educating humans. Building artifacts in space encompasses diverse activities such as in-situ resource utilization, design, manufacturing, verification, validation, and testing, and self-replication. 

A summary of the AI use cases for space exploration are presented in Table \ref{table:Table1}.

\begin{table}
\centering
\begin{tabu} to 0.8\textwidth { | X[l] | X[l] | }
 \hline
 \textbf{AI use case} &  \textbf{Reference} \\
 \hline
 \textbf{\textit{Current and near-term}}   &  \\
\hline
Detect and characterize features of interest (usual / unusual; static/dynamic) & \cite{Chien2006,Chien2017} \\
\hline
autonomous collection of interesting samples & \cite{Chien2006,Chien2017} \\
\hline
autonomous creation of environmental maps & \cite{Chien2006,Chien2017} \\
\hline
on-board analysis of data & \cite{Chien2006,Chien2017} \\
\hline
mission operations planning and scheduling & \cite{Chien2006,Chien2017} \\
\hline
Maintenance (problem analysis, perform repair actions, and evaluate the impact on future operations) & \cite{Bond1978,Chien2006,Hein2012b} \\
\hline
 \textbf{\textit{Mid- and far-term}}   &  \\
\hline
Design and construction of artifacts (spacecraft, infrastructure, colonies) & \cite{Freitas1980a,Freitas1982,Hein2014d,Hirai2014,Jr1981} \\
\hline
In-situ resource utilization & \cite{Freitas1980a,Freitas1982,Hein2014d,Hirai2014,Jr1981} \\
\hline
Self-replication & \cite{Freitas1980a,Freitas1982,Hein2014d,Hirai2014,Jr1981,Tipler1986,Tipler1994,Armstrong2013} \\
\hline
Communication with extraterrestrials & \cite{Baxter2013,Bracewell1960} \\
\hline
Educate humans (transmission of knowledge) & \cite{Crowl2012,Hein2012b} \\
\hline
\end{tabu}
\caption{Current, near-, mid-, and far-term use cases for AI in space exploration}
\label{table:Table1}
\end{table}

The existing literature on AI and AGI in interstellar travel and colonization seems to be limited to high-level concepts and there is a lack of a systematic analysis of the role of AI/AGI and synergies with other technologies. This article addresses these gaps by analyzing the capabilities of AI, AGI for different interstellar missions, explores synergies with other technologies that could result in radically different mission architectures. Concepts for different AI interstellar probes and a generic AI probe design are presented, using the methodology of explorative engineering \cite{Drexler1991,Drexler2013}. 

\section{Analysis Framework} \label{AF}
An analysis framework for AI probes is developed, in order to compare the capabilities required by such probes for fulfilling specific mission objectives. 

Categorizing and measuring capabilities of artificial intelligence is considered challenging and none of the proposed frameworks has been generally accepted \cite{Hernandez-Orallo2017a}. 
Taxonomies categorize artificial intelligence with respect to its abilities (weak vs strong AI; narrow / general AI, super intelligence) \cite{Hernandez-Orallo2017a}, working principles \cite{DeGaris2010, Goertzel2010a}, internal processes \cite{Hintze2016}, embodiment \cite{Goertzel2006}. 
AI metrics are either task-oriented or ability-oriented \cite{Hernandez-Orallo2017a, Hernandez-Orallo2017}. Most existing metrics fall into the task-oriented category, where the performance of an AI system is measured with respect to a task such as playing chess and autonomous driving. Such an evaluation is appropriate for specialized AI systems for specific tasks. By contrast, ability-oriented metrics focus on the set of tasks that would indicate the presence of a more general AI ability, for example, the AI decathlon \cite{ Vere1992, Anderson2003, Hernandez-Orallo2017, Mueller2007, Mueller2008, SimpsonJr2008}. Such an evaluation is appropriate for AI systems that are not characterized by a set of tasks such as cognitive robots, assistants, and artificial pets. 

We propose a mix between more formal and qualitative framework elements. Formal approaches permit the generation of sufficiently general results that might remain valid, even with the large uncertainties associated with future progress in AI. We specifically use the \textit{pragmatic general intelligence} metric \cite{Goertzel2010} for formally comparing different AI capabilities. The qualitative approaches such as literature surveys of both, the scientific literature and fiction, generation of mission architectures, and design of a generic AI probe, allows for exploring specific scenarios and concepts. 

Regarding formal elements of the analysis framework, we argue that any AI-based interstellar mission is based on one or more agents. According to Franklin and Gaesser \cite{Franklin1996}, an agent "is a system situated within and a part of an environment that senses that environment and acts on it, over time, in pursuit of its own agenda and so as to effect what it senses in the future." An agent is distinguished from computer programs in general by their autonomous, adaptive nature. We can define an agent more formally as a function $\pi$ which takes an action history as input and outputs an action. An action history is defined as agent's actions $a$, observations $o$ of the environment, and rewards $r$:

\begin{equation}
a_1o_1r_1a_2o_2r_2...
\end{equation}

The history up to a point in time $t$ can be abbreviated by $aor_{1:t}$.

According to Legg and Hutter \cite{Legg}, the utility function $V$, which expresses the expected total reward $E$ for an agent $\pi$ and environment $\mu$ over its entire lifetime $T$ is: 

\begin{equation}
V^{\pi}_{\mu}\equiv E(\sum^{T}_{n=1}r_1)\leq 1
\end{equation}

Goertzel \cite{Goertzel2010} extends this framework by adding functions that indicate the complexity of goals and environments in which the agent operates, thereby formalizing \textit{pragmatic general intelligence}, defined as achieving complex goals in complex environments. The expected goal-achievement is defined as

\begin{equation}
V^{\pi}_{\mu,g,T}\equiv E(\sum^{t}_{i=s}r_g (I_{g,s,i})\leq 1
\end{equation}

with the interaction sequence $m_1a_1o_1g_1r_1m_2a_2o_2g_2r_2...$, where $m$ is a memory action, and $T={i \in (i,...,t)}$. Each finite interaction sequence $I_{g,s,t} = aorg_{s:t}$ with $g_s$ corresponding to a goal $g$, is mapped by each goal function to a 'raw reward' $r_g(I_{g,s,t}) \in [0,1]$, indicating the reward of achieving the goal during that interaction sequence. The agent's total reward $r_t$ is the sum of the raw rewards from all goals obtained at time $t$, where the symbols for these goals appear in the agent's history before $t$.

According to Goertzel \cite{Goertzel2010}, the \textit{pragmatic general intelligence} of  an  agent $\pi$,  relative  to  the  distribution $\nu$ over  environments  and  the  distribution $\gamma$ over  goals,  is  its  expected performance  with  respect  to  goals  drawn  from $\gamma$ in environments drawn from $\nu$ 

\begin{equation}
\Pi(\pi)\equiv \sum_{\mu\in E, g\in \mathcal{G}, T} \nu(\mu) \gamma(g,\mu) V^{\pi}_{\mu,g,T}
\end{equation}

This formal framework of pragmatic general intelligence allows for a comparison of the intelligence of agents as a sum of the expected rewards these agents would obtain with respect to environments and goals. For example, an agent that is expected to obtain a reward 0.2 in a single environment $\mu$ with respect to goals $g_1$ and $g_2$ has a total $\Pi$ of 0.4, whereas an agent that is expected to obtain a reward 0.4 for goal $g_1$ in a single environment $\mu$ has the same value for of 0.4. Hence, the metric allows for taking both, breadth and specificity of an agent's performance into account. 

Apart from this quantitative framework for comparing an agent's capability, we further use a qualitative maturity scale for analyzing task-specific capabilities and general capabilities with respect to AI probe missions, drawing heavily from Hernandez-Orallo \cite{Hernandez-Orallo2017,Hernandez-Orallo2017a} and Hein \cite{Hein2016c}. The results of this qualitative analysis are presented in Section 4.

\section{Artificial intelligence probe concepts}
AI probes can be distinguished with respect to their objectives. It could be classic exploration where AI serves only as a means for realizing autonomous exploration of a star system. It could also be more sophisticated such as preparing an infrastructure for human colonization or even an entirely AI-based colonization. We distinguish between four types of AI probes: 

\textit{Explorer –}
\begin{itemize}
\item capable of implementing a previously defined science mission in a system with known properties (for instance after remote observation); \cite{Baxter2013,Hein2011}
\item capable of manufacturing predefined spare parts and components;
\item Examples – the Icarus and Daedalus studies \cite{Bond1978}. 
\end{itemize}

\textit{Philosopher –}
\begin{itemize}
\item capable of devising and implementing a science program in unexplored circumstances; 
\item capable of original science: observing unexpected phenomena, drawing up hypotheses and testing them; 
\item capable of doing this within philosophical parameters such as planetary protection; 
\item capable of using local resources to a limited extent, e.g. manufacturing sub-probes, or replicas for further exploration at other stars.
\end{itemize}

\textit{Founder –}
\begin{itemize}
\item capable of using local resources on a significant scale, such as for establishing a human-ready habitat; 
\item capable of setting up a human-ready habitat on a target object such as part of an embryo space colonization programme;
\item perhaps modifying conditions on a global scale (terraforming). \cite{Fogg1991,Hein2014d,Hein2012b}.
\end{itemize}

\textit{Ambassador –}
\begin{itemize}
\item equipped to handle the first contact with extraterrestrial intelligence on behalf of mankind, within philosophical and other parameters: e.g. obeying a Prime Directive and ensuring the safety of humanity. \cite{Baxter2013,Bracewell1960}
\end{itemize}

A more detailed description of each of the probe types is given in the following. Besides the scientific literature, the science fiction literature has elaborated on different probe types and will be considered. 

\subsection{Explorer}
An Explorer probe is an extension of the model of modern-day automated space probes, which have limited on-board AI and well-defined missions. 
Because of remoteness from Earth modern probes are capable of some independent decision-making. Probes may put themselves into ‘safe' modes in case of navigation failures or other issues; Mars rovers will stop before or back up from unexpected obstacles. But essentially, in the event of novelty, the probes wait for further orders from an Earth-bound mission control. 

Because of light speed time delays, this would not be an option with a probe like Icarus and Dragonfly to Alpha Centauri \cite{Hein2017,Hein2016b,Long2009,Lubin2016,Perakis2016}. During the long flight, the AI would need to deal with routine systems operations like course corrections and communications, and also maintenance, upgrades, and dealing with unplanned incidents like faults. On arrival at Alpha Centauri, coming in from out of the plane of a double-star system, a complex orbital insertion sequence would be needed, followed by the deployment of subprobes and a coordination of communication with Earth \cite{Baxter2016a}. It can be anticipated that the target bodies will have been well characterised by remote inspection before the launch of the mission, and so objectives will be specific and detailed. Still, some local decision-making will be needed in terms of handling unanticipated conditions, equipment failures, and indeed in prioritising the requirements (such as communications) of a multiple-subprobe exploration.

\subsubsection{Technology and capabilities}
The AI could be based on already existing technologies such as deep learning \cite{LeCun,Schmidhuber2015} for feature recognition and genetic algorithms for task sequencing. Such an AI would not be considered an agent according to the definition of Franklin and Graesser \cite{Franklin1996}, where a distinction is made between programs that just interact with the environment and agents that show a level of autonomy and adaptability with respect to the environment. Using a pre-trained deep learning algorithm would have a limited ability to adapt to a changed environment due to its dependency on large training data sets. The genetic algorithm's performance depends on a carefully crafted set of objective functions, which are hard-wired and not changed during the mission. 

Using the pragmatic general intelligence metric from Section \ref{AF}, we claim that the reward of such an AI for $\mu$ is close to 0 whenever the environment does not resemble the test data set. Keeping things simple, we can define a distribution of environments $\nu_{solsys}$, which represents the distribution of environments within the solar system. We assume that any environment that is sufficiently outside this distribution results in a reward close to 0. Hence, the more $\nu$ differs from $\nu_{solsys}$, $\Pi$ for $\pi_{explorer}$ will approach 0. We can express the similarity of the distribution of environments by a similarity function $sim$ with $sim(\nu_1,\nu_2)=1$ for $\nu_1 = \nu_2$ and  $sim(\nu_1,\nu_2)<1$ for $\nu_1 \neq \nu_2$.With the similarity function approaching 0, the pragmatic general intelligence of the explorer probe would reach 0. 

\begin{equation}
\Pi(\pi_{explorer}) = \lim_{sim(\nu,\nu_{solsys})\to 0} (\sum_{\mu\in E, g\in \mathcal{G}, T} \nu(\mu) \gamma(g,\mu) V^{\pi}_{\mu,g,T}) = 0
\end{equation}

A more advanced Explorer probe could make use of on-board manufacturing capabilities for creating mechanisms, instrument components, and tools for having the flexibility to react to unexpected situations. The existing literature on using manufacturing technologies in space can serve as a source of inspiration \cite{Freeman2017,Jr2017,Skomorohov2016a,Owens2015,Owens2016,Hoyt2016,Trujillo2017}. One possibility is to carry bulk material stocks and a 3D-printer to manufacture components during its trip and during exploration \cite{Freeman2017}. Used components could be recycled to close the material loop once components are no longer needed. Existing and near-term in-space manufacturing technologies rather focus on manufacturing structural elements \cite{Skomorohov2016a} using bulk material sent from Earth, processed in-situ materials on planetary surfaces \cite{Owens2015}, or on small bodies \cite{Dunn2017}. 

On-board manufacturing capabilities would increase the flexibility of the probe, i.e. it would increase the space of potential actions with respect to an observation. An explorer probe without on-board manufacturing would have a set of actions $(a_1,...a_n)$ at its disposal and an explorer probe with on-board manufacturing a set of $(a_1,...a_m)$ actions, where $m>n$. For example, the probe could perform an action to manufacture a larger aperture for its telescope and allow for higher-resolution observations. The following inequality does not hold in general for Explorer probes with on-board manufacturing $\pi_{explorerobm}$ and explorer probes without $\pi_{explorer}$

\begin{equation}
\Pi(\pi_{explorerobm})\geq\Pi(\pi_{explorer})
\label{eq6}
\end{equation}

as there are cases where on-board manufacturing could lead to a globally lower value on pragmatic general intelligence. Imagine the case where manufacturing the aperture leads to a shift in the center of mass of the spacecraft which leads to a higher consumption of fuel, leading to a shorter lifetime of the probe and hence lower performance. 
However, we can imagine a special case where an action $a_i$ enabled by on-board manufacturing substitutes for an action $a_j$. the substitution leads to a higher reward $r_i>r_j$ but has no other effect on the entire history $aor_{1:T}$ of the agent from its first cycle to its end of life $T$. For this case, inequality \ref{eq6} would hold.

However, on-board manufacturing or other means of extending the set of actions would not change the fundamental limitation that the available set of actions is pre-defined via the training data and the system would perform poorly in environments that have no resemblance with the training data. 

The general problem of a machine that constructs something has been treated by von Neumann \cite{VonNeumann1966} and Myhill \cite{Myhill1964} with universal constructor theory. A \textit{universal constructor} is a machine $M_a$ that can construct another machine $M_b$ given an instruction $I$:

\begin{equation}
I_b + M_a \rightarrow M_b
\end{equation}

where "$\rightarrow$" indicates the inputs to a construction process on the left side and the created object on the right side. The simple constructor would be equipped with an initial set of instructions $\mathcal{D}$ to build infrastructure elements, instruments, replacement parts, etc., using given instructions:

\begin{equation}
I_i + M_a \rightarrow M_i, I_i \in \mathcal{D}
\end{equation}

$M_a$ could be a 3D-printer or any other machine for manufacturing. 

\subsubsection{Mission architectures}
\begin{figure}[ht]
\includegraphics[height=4cm]{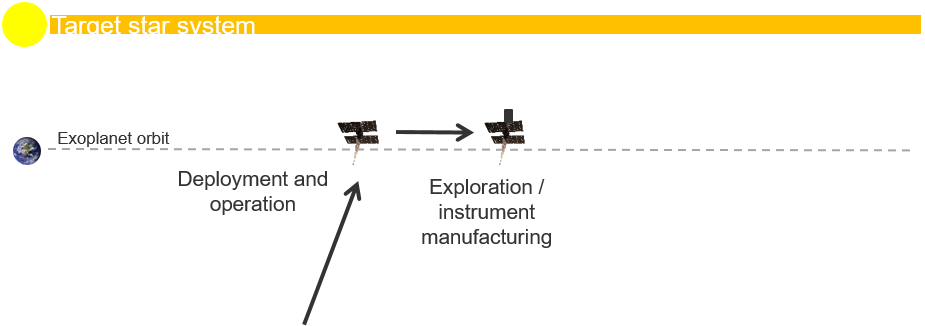}
\centering
\caption{Star system exploration via Explorer probe}
\label{fig:exp1}
\end{figure}

An Explorer type probe would arrive in the target star system and start its exploration program either using its pre-existing hard- and software or could use its capabilities of modifying or manufacturing hard- and software components, depending on the encountered situation. A standard mission architecture for an Explorer type probe is shown in Fig.\ref{fig:exp1}.

\subsection{Philosopher}
In contrast to the Explorers with their specific and well-defined missions, a Philosopher probe is capable of supporting an independent, open-ended exploration strategy. This may include devising and implementing its own science and exploration programme from goal-setting to execution and exploiting local resources to manufacture, for example, subsidiary equipment, subprobes, or even replicas of itself for further interstellar exploration.

\subsubsection{Philosopher probes in fiction}

Philosopher-class probes may be particularly useful on pioneering voyages to develop necessary infrastructure for follow-up missions. In ‘StarCall' \cite{Baxter2016a} a smart probe called Sannah III is sent on an eighty-year mission to Alpha Centauri, using for acceleration a mass-beam propulsion system in Earth orbit, and decelerating using an onboard inertial-confinement fusion drive. Once at Centauri, the mission is to construct another mass-beam propulsion station from local resources; future probes, with no need to carry fuel for deceleration, will be capable of delivering cargoes orders of magnitudes larger.

An interesting advanced probe of the Philosopher type, depicted in Greg Bear's 1990 novel Queen of Angels \cite{Bear1990}, is AXIS (for Automated eXplorer of Interstellar Space), a probe to Alpha Centauri. An advanced onboard ‘biologic thinker system' (Chapter 4) can design its own science programmes at the target. As an example, AXIS observes circular structures on a planetary surface which it hypothesises are artefacts of ETI, a hypothesis it later disproves. 

An example of a wider Philosopher-class probe strategy is Tipler's \cite{Tipler1980} suggestion that the use of self-replicating von Neumann machines \cite{VonNeumann1966} as probes could reduce the costs of a large-scale interstellar exploration programme drastically – the originating culture need only bear the costs of sending out the first probe, and allow descendants constructed of local resources to explore the Galaxy step by step. As a near-term example of this idea, Freitas \cite{Freitas1980a} described a Daedalus probe with a self-replicating payload to establish an industrial infrastructure that allows for building another Daedalus self-replicating probe. However, Freitas did not provide details about the AI required for this task. 

\subsubsection{Technology and capabilities}

One key aspect of the Philosopher probe will be how the AI can learn from the data at the target star system to adapt and learn from new findings. It is quite obvious that optimal problem solvers such as practical implementations of AIXI \cite{Hutter2004,Legg,Veness,Veness2011} and the self-referential Gödel machine could be used \cite{Steunebrink,Steunebrink2012}. Another possibility is to use genetic algorithms to automatically generate programs adapted to new findings at the star system \cite{Becker2017}. 

To take the Gödel machine as an example, it consists of two parts. The first part is a program that interacts with its environment. The second part includes a proof searcher that searches for proofs that a modification to the Gödel machine is expected to yield higher rewards during its lifetime. Once such a proof is found, the modification is implemented and the Gödel machine modified. Schmidhuber \cite{Schmidhuber2009} argues via his Global Optimality theorem that the Gödel machine performs optimally in the set of environments $\nu$ and is not restricted by the Free Lunch theorem. The Gödel machine can modify any part of its code, including the proof searcher itself and the utility function which sums up the rewards. Hence, a Philosopher probe based on a Gödel machine would, in principle, not be bound by the limitations of the Explorer probe and could modify its soft- and hardware with respect to a specific environment and even set its goals. A version of the Gödel machine for solving design problems has been proposed by Hein and Condat \cite{Hein2018d}. Such a design Gödel machine could be used for building infrastructures in the target star system. 

The Gödel machine only switches to a modified version if it can prove that it would yield better results on the utility function, which means that:

\begin{equation}
(V^{\pi_{gm}}_{\mu,g,T}\geq V^{\pi_{explorer}}_{\mu,g,T}),
\forall \nu(\mu)\in E, \forall \gamma(g,\mu) \in \mathcal{G} 
\end{equation}

which means that the Philosopher AI based on a Gödel machine has a practical general intelligence, which is equal or larger that of the Explorer:

\begin{equation}
\Pi(\pi_{gm})\geq\Pi(\pi_{explorer})
\end{equation}

The Philosopher AI would not only be able to manufacture artifacts as the Explorer but go a step further in actually \textit{designing} artifacts during its mission. Using the notation from von Neumann \cite{VonNeumann1966} and Myhill \cite{Myhill1964}, a designing machine is a machine $M_c$ that creates instructions $I_i$: 

\begin{equation}
M_c \rightarrow I_i
\end{equation}

Programs that can synthesize designs are numerous and already exist today, for example, for generating complex geometric shapes from geometric primitives \cite{Chakrabarti2011}. For complex soft- and hardware, a design Gödel machine \cite{Hein2018d} could be imagined, where the machine analyzes its environment (available resources) and a set of design requirements (provide an air-tight volume), synthesizes a set of designs (aluminum hull) and assesses its feasibility. Feasibility is assessed via simulations or testing. We can imagine that the machine conceives small prototypes to test key feasibility areas with minimal expenditure of available resources before embarking to the construction of the real artifact. 

An open research question concerning self-improving AI is the safety problem \cite{Everitt2018}. If the AI can modify itself and specifically its utility function, how can we assure that it will not take harmful actions?

\subsubsection{Self-replicating probes} \label{self-replicating probes}

Self-replicating probes have been proposed in the literature for decades \cite{Freitas1980a,Jr1981,Freitas1982,Metzger2016}. In the following, we are referring to the theory of self-replicating machines which can be found in the theoretical computer science literature such as \cite{Arbib1988,Myhill1964} and \cite{VonNeumann1966}. 

A self-replicating machine can be understood as a machine \(M\), e.g. a Turing machine or equivalent that accepts some input data \(I\) that includes a description of \(M\) and is able to construct \(M\). However, this is not self-replication, as the instruction is not copied. Hence, we introduce two machines \(M_a\) and \(M_b\) where \(M_a\) creates a copy of \(I\) and \(M_b\) creates a copy of \(M_a\) and \(M_b\). The input data \(I\) needs to include a description of \(M_a\) and \(M_b\). 
\begin{equation}
I_{a+b}+M_a\rightarrow I_{a+b}
\end{equation}
\begin{equation}
I_{a+b}+M_b\rightarrow M_a+M_b
\end{equation}
Where "\( \rightarrow \)" can be interpreted as "creates". Hence, combining the two yields a self-replicating machine:

\begin{equation}
I_{a+b}+M_a+M_b\rightarrow I_{a+b}+M_a+M_b
\end{equation}
Although the existence of a self-replicating machine has been formally proven, an actual construction turned out to be more difficult. Programs that can take their own code as inputs have been around for years and are called "Quines" \cite{Hofstadter1979}. Self-replicating machines based on cellular automata have been developed but turn out to be computationally very expensive, as they simulate the assembly or the machine from elementary parts \cite{Sipper1998}. Robotic self-replicating machines have been proposed by Zykov et al. \cite{Zykov2005,Zykov2007}, Yim et al. \cite{Yim}, and Griffith et al. \cite{Griffith2005}. However, they use prefabricated parts that are assembled to form copies of themselves. Several NASA NIAC studies \cite{Lipson2002,Chirikjian2004,Boston2004,Toth-Fejel2004}  have concluded that at least "cranking" self-replicating machines are feasible. Nevertheless, for any practically useful application, physical self-replicating machines would need to possess considerable computing power and highly sophisticated manufacturing capabilities, such as described in Freitas \cite{Freitas1980a,Jr1981,Freitas1982}, involving a whole self-replication infrastructure. Hence, the remaining engineering challenges are still considerable. Possible solutions to some of the challenges may include partial self-replication, where complete self-replication is achieved gradually when the infrastructure is build up \cite{Metzger2012}, the development of generic mining and manufacturing processes, applicable to replicating a wide range of components, and automation of individual steps in the replication process as well as supply chain coordination.

The main challenge for the AI of such a probe is rather how to adapt the design of the probe to the given resources in a star system. Depending on the chemical composition and reachability of resources in the star system, different mining and manufacturing processes are needed. For example, resources on asteroids, comets, exoplanets, and exomoons might be quite different in composition and ease of mining them \cite{Kennedy2015,Beichman2005,Martin2012}. Using mining and manufacturing processes that are applicable to a broad range of resources would significantly facilitate the challenge, e.g. sintering can be applied to a broad range of regolith material, whereas high purity metals and alloys require highly specialized processes, which are limited to a specific type of metal and alloy. However, products from general processes might suffer from lower performance characteristics compared to products of a highly specialized process, e.g. tensile strength. 

A special case of self-replicating machines are self-replicating machines that improve on each generation. Myhill \cite{Myhill1964} provides an existence proof for such machines with the properties: 

\begin{equation}
M_{z_i}<M_{z_{i+1}}, i=0,1,2...
\end{equation}

where "$<$" indicates that the machine on the right has larger theorem-proving capabilities than the machine on the left and $i$ indicates the generation and 

\begin{equation}
M_{z_i}\rightarrow M_{z_{i+1}}
\end{equation}

where a machine produces a machine with greater capabilities of the subsequent generation. 

\subsubsection{Mission architectures}

\textit{Surface exploration, including astrobiology}

\begin{figure}[ht]
\includegraphics[height=5cm]{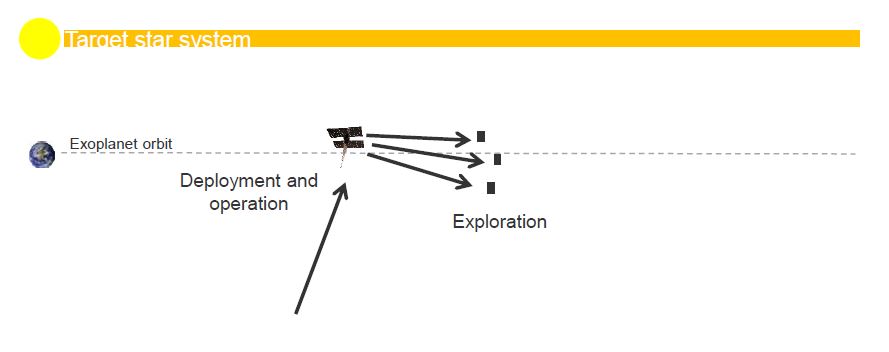}
\centering
\caption{Star system exploration via multiple sub-probes}
\label{fig:phil1}
\end{figure}

Similar to an Explorer type probe, the most basic mission architecture for a Philosopher type probe would consist of the probe arriving in the star system, as shown in Fig. \ref{fig:phil1} and deploying a number of sub-probes for exploration. The difference to an Explorer probe is that the exploration strategy is developed in-situ, depending on the observations the probe will make. 

\textit{Self-replicating star system exploration}

A mass-efficient exploration strategy would comprise the production of self-replicating probes using in-situ resources of the star system. Only the mass for the initial spacecraft is required, thereby exponentially reducing the required mass for exploration. The success of such an exploration strategy will depend on the ease of identification, reachability, and extraction of resources in the star system. Various AI architectures can be imagined. Computing on the sub-probes could be limited and the main probe would be responsible for sophisticated computations. Such an architecture might be more efficient but also more risky, in case of the failure of the main probe. Alternative architectures could be based on distributed computation between sub-probes and the main probe, where the computing power of sub-probes and the main probe would not differ significantly, which would increase the reliability of the overall system. However, such an architecture would require the replication of computing hardware in-situ, which might be difficult to achieve. 

\begin{figure}[ht]
\includegraphics[height=5cm]{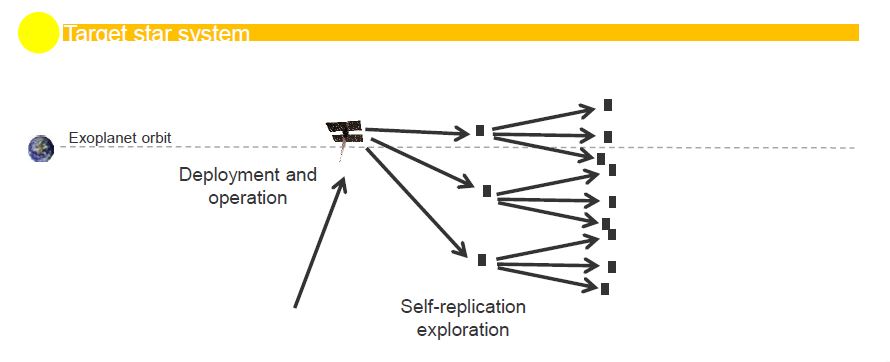}
\centering
\caption{Star system exploration via self-replication}
\label{fig:phil2}
\end{figure}

\textit{Adapted biome creation}

\begin{figure}[ht]
\includegraphics[height=5cm]{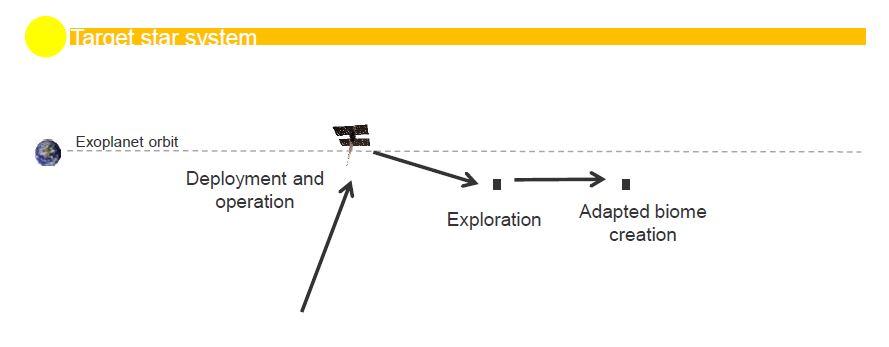}
\centering
\caption{Creation of adapted micro biome for future colonization}
\label{fig:phil3}
\end{figure}

Another mission architecture for a Philosopher probe could consist of the preparation of habitable planets for subsequent settlement. A crucial element for human habitability is the existence of a human-compatible biome, i.e. microorganisms \cite{Davies2013}, which is vital for human survival. In case the exoplanet is sterile, such a biome could be engineered, taking the local environmental conditions into consideration. For example, a higher level of stellar radiation might lead to different environmental pressures on the biome than on Earth, leading to a biome which is no longer sufficient for human survival. Engineering and cultivating an adequate biome would be a task which would require sophisticated AI capabilities. The corresponding mission architecture is shown in Fig. \ref{fig:phil3}

\textit{World model creation}

\begin{figure}[ht]
\includegraphics[height=5cm]{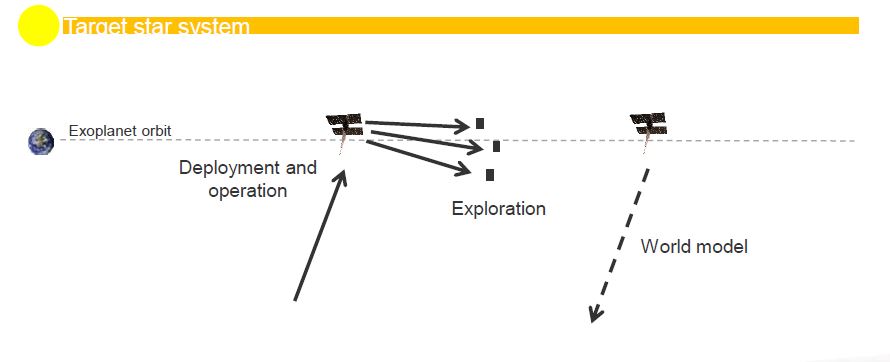}
\centering
\caption{Creation and submission of world model}
\label{fig:phil4}
\end{figure}

Another possible Philosopher probe objective could be the generation of a "world model", as shown in Fig. \ref{fig:phil4}. A world model \cite{Ha2018} in the context of AI is similar to a mental model in humans in which they can perform reasoning without directly taking action on the real world. Here, we can think of world models as either such AI mental models that are transmitted back from the star system to the solar system and could allow for running experiments and simulations. In other words, world models could be used for virtually exploring the star system. 

\textit{Traveling AI}

\begin{figure}[ht]
\includegraphics[height=5cm]{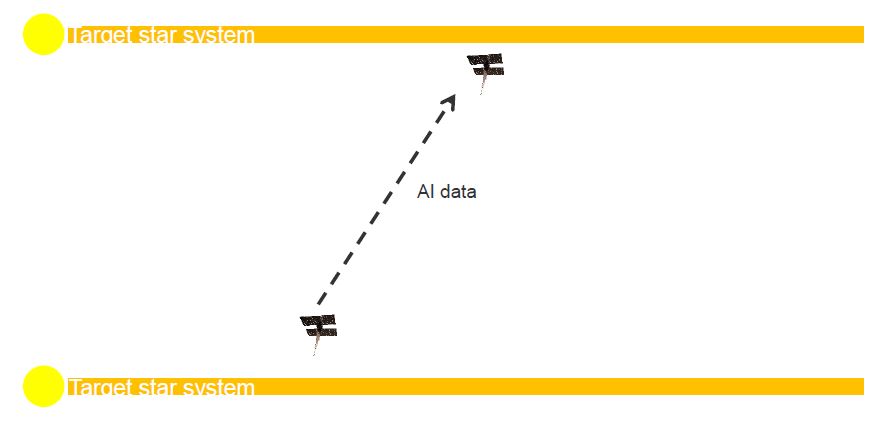}
\centering
\caption{Traveling AI between probes in different star systems}
\label{fig:phil5}
\end{figure}

Another possible mission architecture consists of a traveling AI, as shown in Fig. \ref{fig:phil5}. Once the target star system has been explored, the AI which has been interacting with this environment could transmit a copy of itself back to the solar system. Alternatively, an AI could be transmitted to the star system. This would be interesting, in case the evolution of AI in the solar system is advancing quickly and updating the on-board AI would lead to performance improvements. Updating on-board software on spacecraft is already a reality today \cite{Erickson2006,Lutz2011}. Going even a step further, one can imagine a traveling AI which is sent to the star system, makes its observations, interacts with the environment, and is then sent back to our solar system or even to another Philosopher probe in a different star system. An AI agent could thereby travel between stars at light speed and gradually adapt to the exploration of different exosolar environments. 

\subsection{Founder}
A Founder probe is capable of much more ambitious missions, including the significant modification of its environment, and perhaps even establishing human colonies. Not only does the Founder need the capability to collect and analyze data such as the Philosopher but also the capability to deliberately alter its environment and to verify that the conceived interventions and designs actually work out. Hence, sophisticated simulation, optimization, and reasoning capabilities are required.

\subsubsection{Founder probes in the literature}

The classic application of a Founder-class probe may be the ‘seedship' colony strategy. Crowl et al. \cite{Crowl2012} gave a recent sketch of possibilities for ‘embryo space colonisation' (ESC). The purpose is to overcome the bottleneck costs of distance, mass, and energy associated with crewed interstellar voyages. Crowl et al. \cite{Crowl2012} suggested near-term strategies using frozen embryos, and more advanced options using artificial storage of genetic data and matter-printing of colonists' bodies, and even ‘pantropy', the pre-conception adaptation of the human form to local conditions. Hein \cite{Hein2014d} previously explored the possibility of using AI probes for downloading data from probes into assemblers that could recreate the colonists. Although appearing speculative, Boles et al. \cite{Boles} have recently demonstrated the production of genetic code from data. 

A seedship's AI would, at a minimum, need to create a human-ready habitat from local materials – though more advanced options up to terraforming could be considered. And, crucially, it must raise the first generation of colonists to adulthood, and perhaps beyond, without adult-human support.

\subsubsection{Founder probes in fiction}

In fiction embryo/genetic space colonization had been hinted at as long ago as 1930, by Stapledon in Last and First Men \cite{Stapledon1968}. Some billions of years in the future, the Last Men on Neptune, threatened by solar destabilization (p.238), disseminate ‘among the stars the seeds of a new humanity'. These will ‘combine to form spores of a new life, and [will] develop, not into human beings, but into lowly organisms with a definite evolutionary bias toward the essentials of human nature'. (Earlier in this saga the genetic adaptation of human species to new environments, on Venus and Neptune, was also depicted.)

Clarke's 1986 novel The Songs of Distant Earth \cite{Clarke1986} contains a classic modern description of seedship colonization, with a typically elegant summary of its challenges. Driven by the impending nova explosion of the sun, a ‘seedship' carrying ‘gene patterns' (Chapter 2) colonized an Earthlike planet called Thalassa, fifty light years from Earth. The first generation of Thalassans was manufactured and raised by machines. The ship had to ‘rear these potential humans, and teach them how to survive in an unknown but probably hostile environment. It would be useless – indeed, cruel – to decant naked, ignorant children on to worlds as unfriendly as the Sahara or the Antarctic. They had to be educated, given tools, shown how to locate and use local resources. After it had landed and became a Mother Ship, it might have to cherish its brood for generations. Not only humans had to be carried, but a complete biota.' (p13). Seven hundred years later, ‘the Mother Ship [was] the oldest and most revered monument on the planet' (p14).
Vinge's ‘Long Shot' (1972) \cite{Vinge1972}, perhaps more realistically, hints at the challenges posed even by the journey component of such a mission. When the Earth is threatened by a lethal increase in solar luminosity, a 10,000-year embryo space colonization mission to Alpha Centauri is hurriedly mounted as a last-resort species survival option. The story is told from the point of view of the onboard AI, called Ilse. Ilse is trained in Earth orbit in such disciplines as orbital manoeuvres and planetary survey, manages the long mission itself, observes the target stars and selects a planet for landing, and survives a final drastic atmospheric re-entry. 

But during the long journey component failures degrade Ilse's mentation and memory, to the extent that she struggles to complete her tasks, and even forgets the primary mission. A last-resort backup memory enables the nurturing of the embryos to go ahead – and this convincing story tantalisingly ends before the AI's next great challenge: raising the first generation of colonists. 
Crowl et al (perhaps optimistically) suggested it would be sufficient to use androids as surrogate parents: AIs embodied to enable physical contact, and equipped with ‘a type of expertly programmed expert system, with sophisticated natural language abilities'. We may, however, need a more complete understanding of what contribution other human beings make to our development from infancy before we can be sure how to supplant natural parenting with surrogates. 

This contribution may even include a biological input. The seedship would need the capability of synthesising far more than ‘human' cells. The human body contains ten thousand times as many microbes of specialised kinds as eukaryotic human cells; together this ‘human microbiome' has a gene set far larger than the human. Furthermore, the development of the microbiome in an individual's body is not well understood; perhaps the microbes are transmitted from others, like infections \cite{Scharf2014} (pp142-3).

Cultural learning would also need to be assured. As an extreme case study, Kemp \cite{Kemp2015} speculated on how a group of infants, isolated from any adult contact at all, might develop. Is culture hard-wired into our consciousness? In one relevant example, a sign language spontaneously developed among an isolated group of deaf children in Nicaragua in the 1970s. Necessary if primitive tools might be invented from scratch, and a lifestyle equivalent to hunter-gathering might emerge, depending on the environment. Sexual differentiation of behaviour and roles might arise when the first wave of pregnancies occurred – and the first deaths might lead to religious impulses. Groups limited by the ‘Dunbar number' of ~150 close personal contacts might emerge, leading to differentiation of culture, perhaps even war. 

Clearly, the contribution of the wider environment of our human society to our development will have to be well understood and replicated if humans are to be manufactured ‘from scratch' beginning with nothing but genetic data. Perhaps relevant case studies such as that of the Nicaraguan deaf children could give an indication as to the minimum support, physical and psychological, required of any AI surrogate-parent to raise successfully any seedship children. However, Kemp quotes paleoanthropologist Ian Tattersall as predicting that the stranded group would die out, with the children developing pathologically without the presence of adults: that is, deprived of the social nurturing with which we have evolved. 
Ethical questions also arise. It would presumably be possible to imprint the infants with cultural values of a specific type, such as religious or libertarian. Would it be right to do so? After all, such values are transmitted within any ‘normal' human society from parents and teachers to children. 

On the other hand, a weaker purposeful conditioning may cede unanticipated influence to the unusual initial conditions of the seedship colonists. In Hogan's Voyage from Yesteryear (1982) \cite{Hogan1982}, a limited nuclear war in 1992 triggers a panicky attempt to seed Chiron, a planet of Alpha Centauri. Generations later, humanoid robots with an essentially nurturing role continue to permeate Chironian society – just as during the upbringing of the first generation (p125). Just as there was no obvious hierarchy of human authority presented to the first cadre of children in an adult-free world, they have developed a society which continues to be hierarchy-free and self-organising (p110). And, still more profoundly, the Chironians have continued to regard material goods as free and as abundant as they were when provided by the founding robots in the beginning: ‘the idea of restricting the supply of anything never occurred to anybody. There wasn't any reason to. We've carried on that way ever since. You'll get used to it' (p129). Thus they have naturally evolved a post-scarcity society. 

When more conventional follow-up missions are sent to Centauri, Hogan hints at still deeper cultural clashes between ‘normal' folk and the seed-grown: ‘"It's not really their fault because [the seed-born are] not really people like us ..."' (p44). There may even be religious prejudices. The ship carries an ordinance proclaiming that the seed-born have souls – rather as the sixteenth-century Popes had to decree that the inhabitants of the New World had souls, like Europeans (p44).

In terms of more advanced seeding technologies, the term ‘pantropy' \cite{SFE2015}, meaning the pre-conception manipulation of human stock to adapt it drastically for survival in novel environments, was coined by SF writer James Blish in his ‘Seedling Stars' stories (1952-6, coll. 1957) \cite{Blish1957} (though Blish acknowledged Stapledon's prior exploration of related ideas). If seedships reduce the cost of the travel to a new home, pantropy should reduce the cost of adaptation in a new environment – pantropy, changing people, will be cheaper than terraforming, changing worlds. Blish's story starts with a rogue pantropist who has adapted humans to survive on Ganymede; ultimately a relatively low-cost, open-ended interstellar ‘seeding programme' (p54) succeeds so well that interstellar colonists return to create pantropes to recolonise ‘the vast and tumbled desert of the Earth' (p192).

It would appear that on Clarke's Thalassa \cite{Clarke1986} the education of the initial generations and cultural transmission from the terrestrial precursor went well; the Thalassans understand where they came from, how they got there, and the meaning of later visitors from Earth. But this transmission is a challenge, and perhaps more so if the pantropes' own physical form is drastically modified. A cautionary tale of cultural discontinuity and amnesia is Blish's story ‘Surface Tension' \cite{Blish1952}, in which a crash-landed seedship crew on Hydrot, world of Tau Ceti, hastily creates microscopic pantropes to share mud-flat puddles with a menagerie of algae, diatoms, protozoans, and rotifers. With no knowledge of their origin, even of their basic cosmological context, after sixteen generations the pantropes try to break out of a puddle-world into which they don't seem to fit, delivering the mother of all science-fictional conceptual breakthroughs: ‘the two-inch wooden spaceship and its microscopic cargo toiled down the slope towards the drying little rivulet' (p175) \cite{Blish1957}.

The embryo space colonization idea remains imaginatively alive. In the recent movie Interstellar (2014, dir. C. Nolan), with the Earth becoming uninhabitable due to a blight, embryo space colonization through a wormhole was presented as a ‘Plan B' to save mankind if a ‘Plan A', involving the transport of mature humans, failed.

\subsubsection{Technology and capabilities}
Founder probes will certainly use some form of self-replication, which has been presented in Section \ref{self-replicating probes} in order to bootstrap the infrastructure needed for building up habitats (free-floating or surface colonies, or even terraforming). However, we can expect that the breadth of tasks required for building a habitat is much larger and their safety-criticality much higher. It is also reasonable to assume that the complexity of a space colony is higher than that of a Philosopher probe, for the simple reason that a space colony would likely also contain a sophisticated AI for environmental control and maintenance \cite{Hein2012b}. Furthermore, engineering a proper biome for the ecosystem and fine-tuning the overall system to the local conditions is a task that would be challenging for human engineers. 

Hence, it is reasonable to assume that the pragmatic general intelligence of the AI of a Founder probe is equal or higher than for a philosopher probe, given the larger number of goals to be achieved and higher required performance on these goals ($r_g$ yields values close to 1 only if a number of strict safety conditions are satisfied). Hence, the \textit{expected goal-achievement} for the Founder is equal or higher than for the Philosopher:

\begin{equation}
V^{\pi_{founder}}_{\mu,g,T}\geq V^{\pi_{philosopher}}_{\mu,g,T}
\end{equation}

If we assume that the distributions $\nu$ over environments and over goals $\gamma$ are the same as for the Philosopher probe, we yield

\begin{equation}
\Pi(\pi_{founder})\geq\Pi(\pi_{philosopher})
\end{equation}

This does not exclude that the Philosopher probe AI could achieve higher raw rewards for individual goals such as devising scientific hypotheses. However, we assume that most of the Philosopher probe's goals are part of the Founder probe's goals. 

\subsubsection{Mission architectures}

\begin{figure}[ht]
\includegraphics[height=5cm]{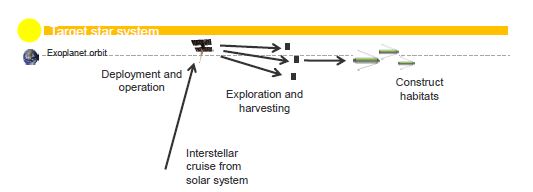}
\centering
\caption{A Founder probe building habitats in a star system}
\label{fig:founder1}
\end{figure}

Fig. \ref{fig:founder1} shows a mission sequence for the Founder probe, which begins with exploring and harvesting the star system to design and manufacture habitats. In-space colonies and surface colonies could be constructed, depending on the judgment of the probe's AI. 

\begin{figure}[ht]
\includegraphics[height=5cm]{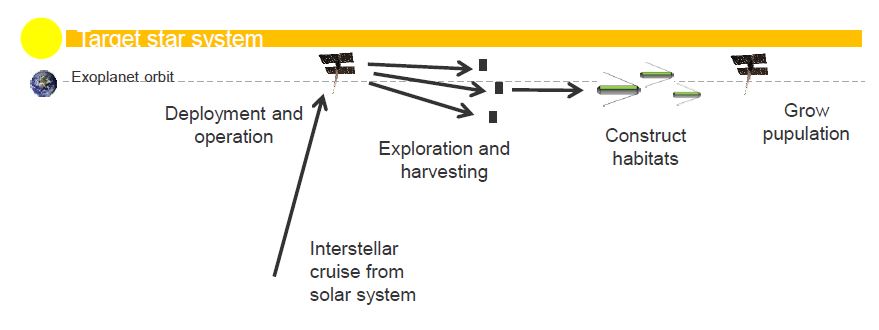}
\centering
\caption{A Founder probe building habitats and growing a population from transported genetic material}
\label{fig:founder2}
\end{figure}

The inhabitants could be transported to the star system via a world ship \cite{Hein2012b,Hein2014d}. Due to the extremely high cost of a world ship, two alternatives can be imagined: The genetic material for creating humans or other organisms is transported via the Founder probe or a separate probe, as shown in Fig. \ref{fig:founder2}. 

\begin{figure}[ht]
\includegraphics[height=5cm]{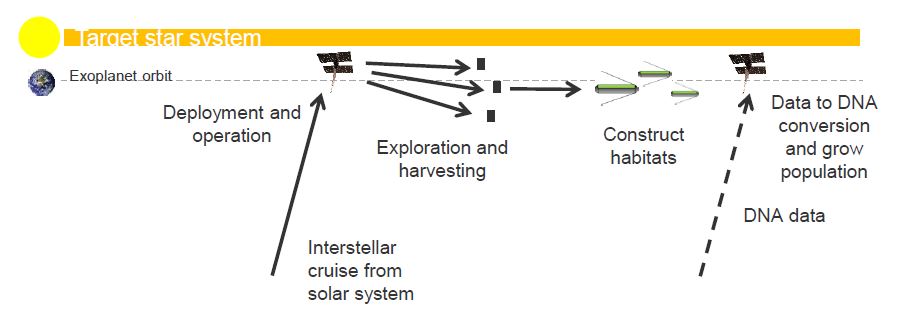}
\centering
\caption{On-site production of genetic material via a data to DNA converter}
\label{fig:founder3}
\end{figure}

As an alternative, the genetic material is recreated from data, using an advanced version of a digital to DNA converter \cite{Boles}. The latter approach would have the advantage that up-to-date DNA data could be transferred at light speed to the star system. One of the caveats of the first instance of a digital to DNA converter is its extremely low efficiency (99.999$\%$) \cite{Pearson2017}. Nevertheless, this would come quite close to the notion of 'teleportation' \cite{Hein2014d}, as illustrated in Fig. \ref{fig:founder3}. 

It is not that outlandish to assume that, for example, stem cells could be transported to the star system, DNA data is sent to the converter, DNA is printed out and the stem cell "reprogrammed". Using far more advanced forms of bio-printing \cite{Murphy2014,Jakab2010} than exist today, entire organisms could be created on-site. Such an approach would circumvent potential radiation-related and age-related degradation during transport in interstellar space.  

We can even imagine that using self-replication technology and advanced manufacturing systems, a design for an up-to-date digital to DNA converter could be sent to the target star system, the converter would be built by the advanced manufacturing system. Hence, a combination of these advanced technologies would allow for significant flexibility in how the colonization operation is performed.  

\subsection{Ambassador}

\subsubsection{Ambassador probes in the literature}

The idea of using smart space probes as a specific means to make contact with extraterrestrial civilizations dates back to Bracewell \cite{Bracewell1960,Bracewell1975}, who proposed the idea in 1960 as an alternative to the then-nascent ‘conventional' SETI model (detection of EM signals). Bracewell imagined a culture sending out many minimal-cost probes equipped with artificial intelligence at least at the human level. On encountering a target culture with radio technology, a probe would initiate contact, perhaps by echoing back native signals. 

This approach has distinct advantages, at least for a long-lived culture, in a universe in which technological cultures are separated by large distances (Bracewell suggests ~1000 light years or more), or, indeed, such cultures are typically short-lived. A local probe would allow rapid dialogue, compared to an exchange of EM signals which might last millennia. The probe might even be able to contact cultures lacking advanced technology, through recognizing surface structures for example \cite{Baxter2016}. And if technological cultures are short-lived, a probe, if robust enough, can simply wait at a target star for a culture ready for contact to emerge – like the Monoliths of Clarke's 2001 \cite{Clarke1968}.
In Bracewell's model, the probe would need to be capable of distinguishing between local signal types, interpreting incoming data, and of achieving dialogue in local languages in printed form – perhaps through the use of an animated dictionary mediated by television exchanges. In terms of message content, perhaps it would discuss advances in science and mathematics with us, or ‘write poetry or discuss philosophy' (p79). 

However, any engagement with an alien culture on behalf of humanity might require a sophisticated political and ethical understanding. Bracewell suggested his probe might need to handle political complications, such as avoiding rivalries between contacted groups by selecting a ‘competent worldwide entity', like Earth's NASA \cite{Bracewell1975} (p79), to speak through. As suggested by Baxter \cite{Baxter2016}, such a probe would presumably need mandates not to harm the extraterrestrial intelligence culture by the violation of a ‘Prime Directive', and not to risk harm to humanity, for instance by revealing the existence of Earth and its location to a potentially hostile culture; it might choose to conceal or mask its approach trajectory, for example. Using such precedents as planetary protection protocols and the First SETI Protocol, before the launch of any such probe a publicly debated and agreed policy on the balance between the opportunities offered by contact and the risks posed by exposure could be developed as a protocol to guide the AI in its decision-making. 

Bracewell went into no details of the probe's AI, beyond speculating that ‘presumably the computing part need only be the size of a human head, which is, we know, large enough to store an immense amount of information' \cite{Baxter2013} (p79). Bracewell's argument was developed further over the years \cite{Freitas1983,Freitas1983a,Freitas1980b,Kuiper1977,Tough1998}. Tarter \cite{Tarter1996} speculated on the use of nanotechnology to send out extremely small smart probes. 

\subsubsection{Ambassador probes in fiction}

In Clarke's 2001: A Space Odyssey (1968) \cite{Clarke1968} the rogue computer Hal was effectively an Ambassador. The true mission of the spacecraft Discovery, to investigate the alien Monolith orbiting Jupiter (in the movie and sequels; Saturn in the novel), was kept secret from the pilot crew of Bowman and Poole, and was known only to a hibernating team of specialists – and to the HAL 9000 unit, the on-board AI. The need to perpetuate this dishonesty caused Hal to break down. But if the crew were incapacitated and contact with Earth lost, Hal himself had been instructed to continue the mission of alien contact (pp.98-9).

Clarke's The Fountains of Paradise (1979) \cite{Clarke1979}, a novel of the building of a space elevator, features a kind of cut-price Bracewell probe, a visitor to the solar system called Starglider by human observers. Arriving in the 2060s, the probe was launched from a red dwarf system 52ly away; proceeding by means of gravity assists it has hopped from star to star, taking 60,000 years to reach Sol. When it arrives in the solar system Starglider initiates conversation using English and Mandarin acquired from our leakage broadcasts. Starglider's primary function seems to be the acquisition and sharing of information: ‘Starglider combines the functions both of Ambassador and Explorer' (p83).

But Starglider may have a wider agenda of cultural manipulation. It reveals ‘almost no advanced technology, and so [has] minimal impact upon the technically-orientated aspects of our culture' (p174). But on the other hand it appears purposefully to demolish religion, for example by logically deconstructing St Thomas Aquinas: as Clarke puts it, Starglider ‘had put an end to the billions of words of pious gibberish with which apparently intelligent men had addled their minds for centuries' (p.94). If intentional, this may amount to a very subtle cultural manipulation (towards a cautious development of technology and away from religion?), implying a deep apprehension of our culture and a very high level of cognition.

\subsubsection{Technology and capabilities}
The capabilities of an Ambassador probe AI are quite distinct from those of previous probe types, as the focus of the former is on communication, i.e. agent to agent interaction. This interaction can be broken down into performing an action (sending a visual signal, moving an object, etc.) and interpreting the action of the other agent. This generic framework would apply to various forms of organisms / ETI. 

An existing formal framework we refer to in the following is multi-agent reinforcement learning \cite{Hu1998,Busoniu2010,Vlassis2007,Tan1993,Littman1994}. It is an extension of reinforcement learning to the multi-agent case with two or more agents \cite{Busoniu2010,Vlassis2007}. 

Using the notation from Busoniu et al. \cite{Busoniu2010}, a single agent reinforcement learning is described via the Markov decision process, where a finite Markov decision process is a tuple $ \langle X, U, f, \rho \rangle $ with $X$ as the finite set of environment states, $U$ as the finite set of agent actions, $f:X \times U \times X \rightarrow [0,1]$ as the state
transition probability function, and $\rho: X \times U \times X \rightarrow R$ as the reward function.

Multi-agent reinforcement learning is a generalization of the single agent case and called \textit{stochastic game}. A stochastic game is a tuple $\langle X,U_1,...,U_n, f,\rho,...,\rho_n \rangle $ with $n$ the number of agents, $X$ the finite set of environment states, $U_i, i = 1,...,n$ the finite sets of actions available to the agents, yielding the joint action set $\textbf{\textit{U}} = U_1 \times...\times U_n, f : X  \times \textbf{\textit{U}} \times X \rightarrow [0,1] $ is the state transition probability function, and $ \rho_i : X \times \textbf{\textit{U}} \times X \rightarrow \mathbb{R}, i = 1,...,n$ are the reward
functions of the agents.

Muti-agent reinforcement learning distinguishes between cases where the agents cooperate, compete, and mixed cases. For the cooperative case, the reward function of the agents are the same ($\rho_i=\rho_j, \forall i,j \in {1,...,n}$). For the competitive case, the reward functions of the agents are distinct ($\rho_i \neq \rho_j, \forall i,j \in {1,...,n}, i \neq j $). The mixed case is neither fully cooperative nor competitive. 

Based on this basic multi-agent framework, we can already draw a few conclusions for an encounter between an Ambassador probe AI and an ETI. Whether or not the ETI is an AI is secondary for applying the formal framework, however, recent publications in the field of Search for Extraterrestrial Intelligence (SETI) have argued for the case of an alien AI \cite{Dick2003,Bradbury2011}. First, it seems very unlikely that the reward functions of these agents are the same, as there is a vast space of possible reward functions and the probability of the agents having the same reward function should be very low unless there is some form of universal convergence. It follows that the interaction between the agents is very likely not cooperative. In case the encounter is between a single Ambassador probe AI and a single ETI, the stochastic game is necessarily competitive if more than one probe AI or ETI are involved, we either have a competitive or mixed case.

Drawing from game theory, the case of an encounter of an Ambassador probe AI with an ETI can be interpreted as the case of coupled learning \cite{Vlassis2007}, where each agent attempts to model the other agent(s), i.e. their transition function(s) and reward function(s). Depending on the type of game, specifically zero-sum, general-sum, or coordination game, different solutions can be calculated by the agents:  Nash equilibrium, correlated equilibrium, or a coordinated joint action. Such solutions cannot always be calculated but have been successfully applied in practice \cite{Vlassis2007}. 

The Ambassador probe AI's actions would correspond to strategies in game theory. We can imagine actions such as 'observe', 'contact', 'send message', 'withdraw', or even 'self-destruct', in case of a hostile ETI. In each time step, the model of the other agent is refined and the next action taken with respect to the model. Regarding the model for the ETI's actions, we face a principal challenge of predicting the actions of an agent that is more powerful than the Ambassador probe AI, which will be addressed in more detail in Section \ref{safety of encounters}.  

Sending back the interaction history of an Ambassador probe's encounter with an ETI could be very useful, as it could form the basis for training future agents to interact with the ETI. Even the transmission of an updated AI to the Ambassador probe could be imagined if the encounter duration might be stretched to decades and longer. 

\subsubsection{Mission architectures}

Mission architectures for the Ambassador probe are likely to resemble those of the Explorer and Philosopher if they are part of an exploration mission. A possible setting is where the Ambassador AI is stored on an Explorer or Philosopher probe and tries to identify cues for ETI in the incoming data. The world model developed by the Explorer / Philosopher AI could also serve as a source for cues for how to communicate with an ETI. 

\subsubsection{Safety of encounters with an alien AI}
\label{safety of encounters}

The basic tasks of an Ambassador probe would be to communicate with an extraterrestrial intelligence, which means first, that it is able to understand signals from such an intelligence, and second, it is able to send signals that can be understood by the intelligence. Such an interaction sequence can be interpreted in the previously introduced agent-based framework. However, a particular challenge is to avoid actions that could be interpreted as hostile or could otherwise have negative consequences. Furthermore, recent SETI / SETA publications have conjectured that an extraterrestrial intelligence might not be biological but an advanced AI itself \cite{Bradbury2011,Dick2003,}. Hence, we have reason to believe that if the Ambassador probe makes contact with an extraterrestrial intelligence, such an intelligence might not be biological in nature but a kind of machine, and more importantly, it might have more advanced AI capabilities than the Ambassador probe itself. 

Intuitively, we would expect that communication with such an advanced AI would be challenging for the Ambassador. We will argue that it is possible, at least for a limited formal case to prove that it is in general impossible to fully interpret the actions of such an AI via an Ambassador probe. One might argue that the formal case, where we refer to theorem proving, is not applicable to a real encounter with an alien AI. However, given that we use formal theorem proving techniques for verifying computer programs that have to adhere to strict safety standards, we still think that such an approach would be suitable for shedding light on some fundamental issues regarding the encounter with alien AI. 

Furthermore, we argue that it is impossible to generally prove that actions the Ambassador would take in interacting with the alien AI are "safe". 

The first case is analogous to the difficulty of "ensuring that the initial agent's reasoning about its future versions is reliable, even if these future versions are far more intelligent than the
current reasoner" \cite{Fallenstein2015}. This type of reasoning has been called \textit{Vingean reflection} in the literature and may apply to humans reasoning about super-human AI as well as AI reasoning about more intelligent versions of itself. Here we argue that the same line of reasoning on Vingean reflection can be applied to the case of an AI on an interstellar probe encountering a more intelligent ETI. Fallenstein and Soares \cite{Fallenstein2015} use backward induction as an illustrative example that an agent which is capable of reasoning about improved versions of itself would need the reasoning capabilities of its improved versions to do so. LaVictoire \cite{LaVictoire2015} uses Löb's theorem to show that an AI's reasoning about a more powerful version of itself is unreliable. Several remedies for this "Löbstacle" have been proposed \cite{Fallenstein2014,Yudkowsky2013,Garrabrant2017a,Garrabrant2016}. 

There are, though, differences to those settings in the literature compared to the case of an encounter with an ETI. Firstly, we have good reasons to believe that such an AI would be vastly superior to an AI we have sent to the stars, as it is likely that such an AI has developed well before we would have developed an advanced AI, therefore having had much more time to evolve. Secondly, the problem is not to predict if modifications made to an AI are potentially harmful but to a certain extent predict the actions of an ETI. Let's suppose that we could have access to the entire code of the ETI. Such a case would happen when we receive a signal from ETI, which might turn out to be a program \cite{Turchin2018}. A program that would be able to prove that such a program is safe would need to be at least as powerful as the program it checks. We simply refer to Löb's theorem for a proof: 

Let $X$ be any logical statement and $L(X)$ be the statement "if ProofSeeker$(X)$ halts, then $X$", where ProofSeeker is a program that searches all possible proofs and halts if and only if one of them is a valid proof of the statement $X$. Löb's theorem states that for all statements $X$, if $L(X)$ is provable, then $X$ is provable.

It is straightforward to apply Löb's theorem to the case of checking whether an alien AI program is safe. In such a case, we take $X$ as the logical statement "alien program is safe". $L(X)$ then translates to "if ProofSeeker("alien program is safe") halts, then "alien program is safe"". The ProofSeeker would be the AI on the Ambassador probe. However, this is in contradiction to the inferior deductive capabilities of ProofSeeker compared to the alien program and therefore, ProofSeeker cannot prove that statement $L("alien \quad program \quad is \quad safe")$. 

It follows that whatever action the Ambassador AI takes, it cannot prove that the ETI would react safely, as it cannot predict that the alien AI's actions would be safe in general. 

The problem of an encounter with alien AI is, therefore, an extreme case of Vingean reflection, where approaches from the literature, such as from Everitt et al. \cite{Everitt2018} (Section 5) that aim at containing potentially harmful self-modifications do not apply. For example, the correct specification of the reward function for avoiding harmful AI does not apply to alien AI, as if there is a reward function, it has already been specified. 

If AI alignment is already a challenge for AI created by humans, it is very likely that ETI is not aligned, neither with human values nor with values of an AI created by humans, e.g. the Ambassador probe's AI, simply due to the vast space of possible AI designs. Unless there is some form of universal convergence of AI designs, it is unlikely that the ETI is similar to the Ambassador's AI. 

Although the values of the Ambassador probe's AI and the ETI are likely misaligned, it might still be possible that the encounter does not result in a harmful result for either side. For example, empathy might be a characteristic that would allow for mutual understanding, where empathy means "the capacity to relate another's emotional state"  \cite{Yalcin2018} and not the general prediction of an agent's actions. Empathy might be linked to an understanding of the other agent's values and adopting these values \cite{Potapov2014}.

To summarize, we have argued that it is in general not possible for an inferior AI to predict all actions of the superior AI and it is likely that the Ambassador probe AI is inferior to an alien AI. Hence, there is no guarantee that we can predict whether or not such an encounter will be safe. It is even more unlikely that the values of these AIs will be aligned, given the vast space of possible AI designs. Nevertheless, characteristics such as empathy, which could still be present, if the AI is able to engage in social interactions, could be a key to mutual understanding.  

\section{Artificial Intelligence Capabilities}
\subsection{Task-oriented capability evaluation}
In the following, we use the task-oriented approach \cite{Hernandez-Orallo2017a,Hernandez-Orallo2017} of comparing the performance of the AI of interstellar probes, looking at the set of tasks the probes need to accomplish. This qualitative approach is in contrast to the more formal approach we have previously taken. The result can be seen in \ref{table:Table3}. 

The Explorer in its basic form has capabilities similar to existing spacecraft for interplanetary exploration. Data collection and processing is performed with large degrees of autonomy. By contrast, the Philosopher is a probe that is able to conduct science autonomously, including devising hypotheses, experimental setups or identifying data collection procedures, and hypothesis testing. We can imagine prototypical forms of such an AI that are based on current machine learning algorithms and a library of scientific hypotheses from which new hypotheses can be derived by recombination and mutation. This includes the identification and analysis of alien life via remote sensing, in-situ analysis of celestial bodies, and analysis of signals \cite{Baxter2013}.

The founder is expected to undertake extensive construction works within the target star system. These could include self-replication, large communication infrastructure with the solar system, space colonies, and even terraforming \cite{Fogg1991,Hein2012b}. The required AI probe capabilities differ. For example, mining resources in-situ, processing, and constructing truss structures is a capability that near-term technology for asteroid mining is likely to be able to accomplish. However, doing so in an environment that is to large extents unknown seems to be much more difficult. Furthermore, the complexity of the systems that are produced influence how sophisticated the AI needs to be. This is due to the emergence of unexpected phenomena in complex systems that require improvisation and creativity to resolve. When it comes to terraforming, the complexity of the system is enormous with limited predictability. Furthermore, there is probably only a limited failure tolerance for such a system. Therefore, complex systems such as space colonies, and very complex systems such as terraforming planets require a broad range of capabilities at similar or superior levels to humans. When it comes to embryo space colonization, capabilities that allow for sophisticated social interactions between the AI and the colonists. 

Finally, the Ambassador has the capability to initiate communication with an extraterrestrial intelligence in addition to the capabilities of the philosopher. Communication requires receiving and decoding signals from extraterrestrials, translating them into a language we are familiar with, composing a message the extraterrestrials are likely to understand, and its transmission. The most critical function is the translation of the signal and the composition of proper responses. Such conversations are imagined by Bracewell (Bracewell, 1960) who assumes a human-level intelligence for the probe. Of course, we can imagine basic conversational capabilities that today's chatbots are already capable of. However, for a low-probability but high-risk event such as contact to an extraterrestrial civilization, we expect that much more sophisticated forms of AI are required that is able to handle the subtleness and ambiguity of language. Another important aspect is social intelligence, including empathy, as we already mentioned in Section \ref{safety of encounters}.

It can be seen in Table \ref{table:Table3} that the Philosopher's set of tasks is a superset of the Explorer's tasks and the Founder's set of tasks is a superset of the Philosopher's. It can also be seen that the Ambassador's set of tasks is a superset of the Philosopher but not a superset of the Founder's set of tasks. 

\subsection{Do we need an AGI?}
We will also briefly touch on the question of whether or not AGI is a precondition for certain AI types. For most applications, no AGI is required, as their mission objectives are related to specific tasks and capabilities for accomplishing them. Hence, narrow but high-performance AI could, in principle, accomplish these tasks. With reference to Hernández-Orallo \cite{Hernandez-Orallo2017a}, general abilities that underlie these specific tasks would be required if the individual tasks would require them, such as those for the Founder probe.

\begin{table}
\centering
\begin{tabu} to 0.9 \textwidth { | X[l] | X[c] | X[c] | X[c] | X[c] |}
 \hline
  &  \textbf{Explorer} & \textbf{Philosopher} & \textbf{Founder} & \textbf{Ambassador} \\
 \hline
 \textit{Image recognition} & X & X & X & X\\
 \hline
 \textit{Hypothesis testing} &  & X & X & X\\
 \hline
  \textit{Signal pattern recognition} & X & X & X & X\\
 \hline
  \textit{Devise scientific hypotheses} & & X & X & X\\
 \hline
  \textit{Universal translation} & & & & X \\
 \hline
  \textit{Conversation} & & & & X \\
 \hline
  \textit{Identify resources} & & & X & \\
 \hline
  \textit{Conceive design (synthesis / analysis)} & & & X & \\
 \hline
   \textit{Resource processing} & & & X & \\
 \hline
   \textit{Construction} & & & X & \\
 \hline
   \textit{Verification, validation, testing} & & & X & \\
\hline
\end{tabu}
\caption{AI probe types and capabilities}
\label{table:Table3}
\end{table}

\subsection{Testing AI capabilities}
Testing AI capabilities prior to an interstellar mission is mandatory. In the following, different test cases for the tasks introduced in the previous section are presented, as shown in Table \ref{table:Table4}. For most of the AI capabilities and tasks, test cases in our solar system or simulated environments can be imagined. 

AI for interstellar probes are likely to be tested for solar system exploration (e.g. Kuiper belt objects, interstellar objects \cite{Heinb}), economic development (e.g. asteroid mining \cite{Hein2018c,Sonter1997,Metzger2016,Metzger2012}, space manufacturing \cite{Skomorohov2016a,Owens2015,Owens2016,Hirai2014}), and colonization (e.g. lunar / Martian base \cite{Eckart1999,Badescu2009,Zubrin2012}, free-floating colonies \cite{ONeill1981,ONeill1977,ONeill1974,Johnson1977,Arora2006}). 

Simulated environments for testing are in widespread use today for testing AI and in robotics. We can also imagine the use of adversarial machine learning and generative adversarial networks (GAN), where agents are pitted against each other, in order to self-train \cite{Huang2011a,Goodfellow2014,Radford2015}. These approaches have recently been used for self-training games \cite{Silver2017} and creating art \cite{Elgammal}.  

Although the solar system provides an environment for testing various AI capabilities, representative conditions under which an AI for an interstellar probe could be tested are more likely to be found in the outer solar system and deep space, due to the signal latency, which makes human intervention difficult. Nevertheless, the solar system environment, supplemented by virtual environments, is likely to be the context in which AI systems are matured before they are sent to the stars.  

\begin{table}
\centering
\begin{tabu} to 0.9 \textwidth { | X[l] | X[l] |}
 \hline
 \textbf{AI capabilities and tasks} &  \textbf{Test cases} \\
 \hline
 \textit{Image recognition} & Simulated images, solar system environment \\
 \hline
 \textit{Hypothesis testing} & Capturing and analyzing data for hypothesis testing in virtual environment, solar system environment \\
 \hline
  \textit{Signal pattern recognition} & Various simulated signals of increased sophistication, test in solar system environment \\
 \hline
  \textit{Devise scientific hypotheses} & Extensive tests for research within the solar system environment\\
 \hline
  \textit{Universal translation} & Decoding language of various forms and of various organisms, including artificially generated languages (e.g. using adversarial machine learning and generative adversarial networks (GAN) where one agent tries to generate new languages that the other cannot translate \cite{Huang2011a,Goodfellow2014,Radford2015})\\
 \hline
  \textit{Conversation} & Training with various living organisms; Training with artificial agents, e.g. agents that are generated to "beat" the AI\\
 \hline
  \textit{Identify resources} & Testing in a solar system environment (e.g. asteroid mining, planetary surface exploration, planetary surface habitat design, space colony construction) and simulated virtual environments \\
 \hline
  \textit{Conceive design (synthesis / analysis)} & Testing in a solar system environment (e.g. asteroid mining, planetary surface exploration, planetary surface habitat design, space colony construction) and simulated virtual environments\\
 \hline
  \textit{Resource processing} & Testing in a solar system environment (e.g. e.g. asteroid mining, planetary surface exploration, planetary surface habitat design, space colony construction \cite{Hirai2014}) and simulated virtual environments \\
 \hline
   \textit{Construction} & Testing in a solar system environment (e.g. e.g. asteroid mining, planetary surface exploration, planetary surface habitat design, space colony construction \cite{Hirai2014}) and simulated virtual environments\\
 \hline
   \textit{Verification, validation, testing} & Testing in a solar system environment (e.g. e.g. asteroid mining, planetary surface exploration, planetary surface habitat design, space colony construction \cite{Hirai2014}) and simulated virtual environments\\
\hline
\end{tabu}
\caption{AI probe capabilities and test cases}
\label{table:Table4}
\end{table}

\section{Design of a Generic Artificial Intelligence Probe}
In the following, we present a concept for an artificial intelligence probe, based on the assumption that any sophisticated AI will still likely use substantial computing resources, thereby consuming substantial amounts of energy. The probe concept has already been presented in Hein \cite{ ArXiv:1612.08733}. In the following, we make the additional assumption that a future human-level or super-human level AI would consume as much energy for its operation as the equivalent energy for simulating a human brain. We think that this assumption is reasonable, given the large uncertainty regarding which path will lead to AGI and as simulating a human brain is considered as one possible pathway towards AGI \cite{Bostrom2014}.

Today's supercomputers use power in the MW range, the human brain, by contrast, uses only about 25 W \cite{Kandel2000}, for a computing power that has been estimated at $10^{20}$ flops. The required power consumption for equivalent computing power using today's or near-future computing hardware can be estimated to be between 1 MW and 100 GW, depending on how far current levels of increase in computing power can be extrapolated into the future. Nevertheless, these figures provide lower and upper bounds for the power requirements to simulate a human brain, i.e. between 10 to $10^{11}$ W. 

The large difference between the lower and upper bounds are also reflected in the scale of the corresponding power generation system in space. For generating power on the order of 10 W, a power generation subsystems for a 3U-CubeSat would be sufficient. For generating $10^{11}$ W, a hundred solar power satellites would be required \cite{Mankins2012}. Furthermore, power generation is likely not required in interstellar space, where power generation is more challenging, due to the absence of stellar radiation. Once the probe arrives in the target star system, we assume that power generation using photovoltaic cells is feasible. 

We argue that heat generation is likely to be a major issue for AI probes, first, due to the large amounts of power consumed, and second, as waste heat generation due to computing is likely to take place within a small volume. We assume that computation is running within a rather small volume, in order to minimize the time of data transfer, as is the case for today's computers. Analogously, we can expect that large amounts of heat are generated in a small volume. Heat rejection is currently a major issue for supercomputers \cite{Nakayama2014} and the predominant approach for heat rejection is the use of heat pipes, which transport a cooling liquid to the processors and the heated liquid away from them. The current heat density in supercomputers is as high as 10 $kW/cm^2$, about one order of magnitude higher than inside a rocket engine nozzle \cite{Nakayama2014}. 

Heat rejection has already been an issue for terrestrial supercomputers, the issue is aggravated in space, where heat can only be rejected without mass loss via radiation, requiring large surface areas facing towards free space. Advanced radiators could reject about $1kW_t/kg$ thermal power per kilogram in the near future \cite{Adams2003,Hyers2012,Juhasz1994}. Consequently, about 100 tons of radiator mass would be required for rejecting 100MW and 1 ton for 1MW respectively.  

One could argue that existing computer architectures are very inefficient in replicating the function of a human brain, resulting in a huge difference in power consumption. Future computer architectures or working principles of computers such as quantum computing could have a disruptive effect on power consumption \cite{Markov2014}. In order not to exclude this possibility, we keep the power consumption of a few dozens to hundreds of Watts as a lower boundary, in case revolutionary new ways are found to reproduce the function of the human brain. However, a more conservative estimate would put the required power at dozens to hundreds of MW for Philosopher, Founder, and Ambassador type probes and lower values for AI that is less sophisticated, such as for an Explorer type probe.  

Regarding the mass of the computing unit, current on-board data handling systems (OBDH) have a computing power on the order of 100 DMIPS per kg. The spacecraft OBDH from the literature have DMIPS values that are about two orders of magnitude below the values for terrestrial processors. If we extrapolate these values to the 2050 timeframe, we can expect spacecraft OBDH with a processing power of 15 million DMIPS per kg. As DMIPS and flops are different performance measures, we use a value for flops per kg from an existing supercomputer (MareNostrum) and extrapolate this value ($0.025*10^{12}$ flops/kg) into the future (2050). By 2050, we assume an improvement of computational power by a factor $10^5$, which yields $0.025*10^{17}$ flops/kg. In order to achieve $10^{20}$ flops, a mass of dozens to a hundred tons is needed. We assume an additional 100 tons for radiator mass and with 1 kW/kg for solar cells, about 100 tons for the solar cells. This yields a total mass for an AI probe on the order of hundreds of tons, which is roughly equivalent to the payload mass of the Daedalus spacecraft of 450 tons \cite{Bond1978}.

Table \ref{table:Table2} shows the mass estimates for the main spacecraft subsystems and its total mass in a 2050 to 2060 time frame. The mass estimate is only valid for the part of the spacecraft that actually arrives at the target star system. 

\begin{table}
\centering
\begin{tabu} to 0.8\textwidth { | X[l] | X[l] | X[l] | }
 \hline
 \textbf{Spacecraft subsystem} &  \textbf{Specific mass} & \textbf{Subsystem mass [t]}  \\
 \hline
 Computing payload & $0.025*10^{17}$ flops/kg & 40\\
 \hline
 Solar cells (current technology) & 1kW/kg & 100\\
 \hline
 Radiators & 1kWt/kg & 100\\
 \hline
 Other subsystems (50 of computing payload) & & 20 \\
 \hline
 \textbf{\textit{Total mass}} & &  \textbf{\textit{260}} \\
\hline
\end{tabu}
\caption{Mass estimate for AI probe in the 2050-2060 time frame}
\label{table:Table2}
\end{table}

Due to the large power consumption and heat rejection requirements, the following characteristics for an AI probe can be concluded:

\begin{itemize}
\item \textit{Large solar panels:} Unless other power sources such as nuclear power is used, the probe will depend on large solar panels / solar concentrators for generating power for the AI payload.
\item \textit{Operation close to target star:} The spacecraft is at least initially operated close to the star. In order to maximize power input from the star and to minimize solar power generator mass, the probe should be located as close to the star as possible. The minimum distance is constrained by the maximum acceptable temperature for the spacecraft subsystems and an eventual heat shield that protects against the starlight. For that purpose, it needs a star-shield to protect against the heat and radiation from the star. A trade-off between the heat shield mass and the mass savings from lower solar power generator mass needs to be made. However, note that the AI payload itself is a source of intense heat and more sensitive spacecraft subsystems such as sensors need to be located distant to the AI payload, e.g. on a boom. 
\item \textit{AI payload switched off outside star system:} AI is switched off outside the target star system, as there is no power source available for its operation. This may also protect against some forms of radiation damage. Nevertheless, proper radiation protection is an issue, as galactic cosmic rays could still destroy circuits via impact and the resulting particle shower;
\item \textit{Large radiators:} A large radiator is needed for rejecting the heat generated by the AI payload; 
\item \textit{Compact computing unit:} The computer is either super-compact or distributed. However, with a distributed system, communication speed becomes an issue and it is therefore likely that the architecture will be as compact as possible to minimize the time for signals to travel within the payload. 
\end{itemize}

\begin{figure}[h]
\includegraphics[height=8cm]{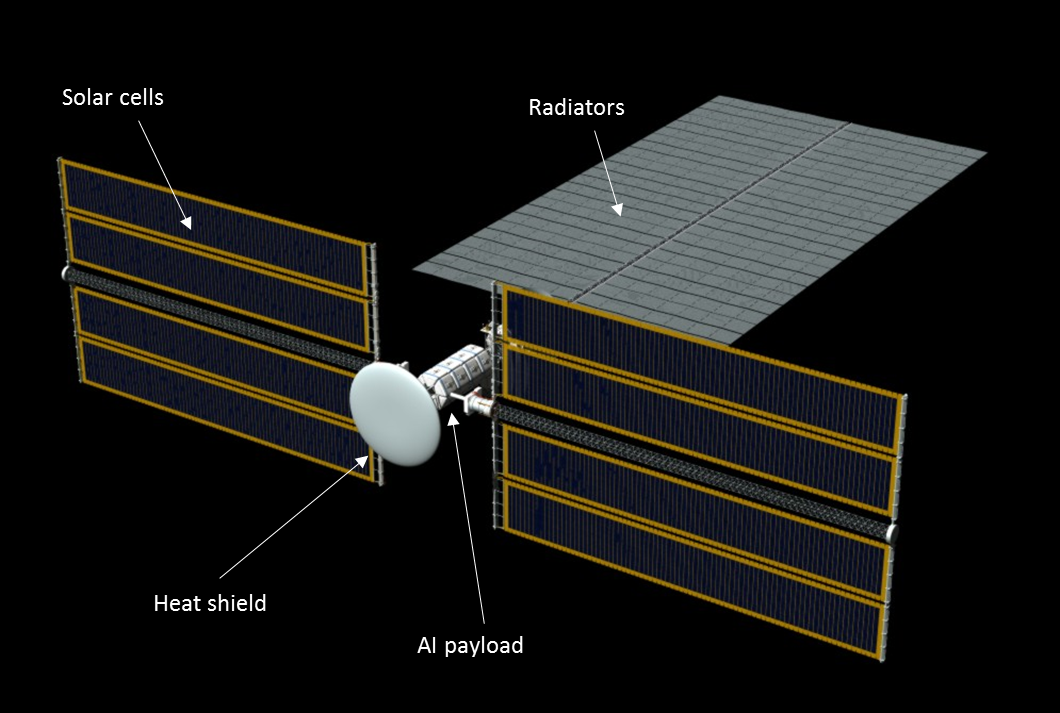}
\centering
\caption{AI probe subsystems (Image: Adrian Mann)}
\label{fig:AIprobe1}
\end{figure}

\begin{figure}[h]
\includegraphics[height=8cm]{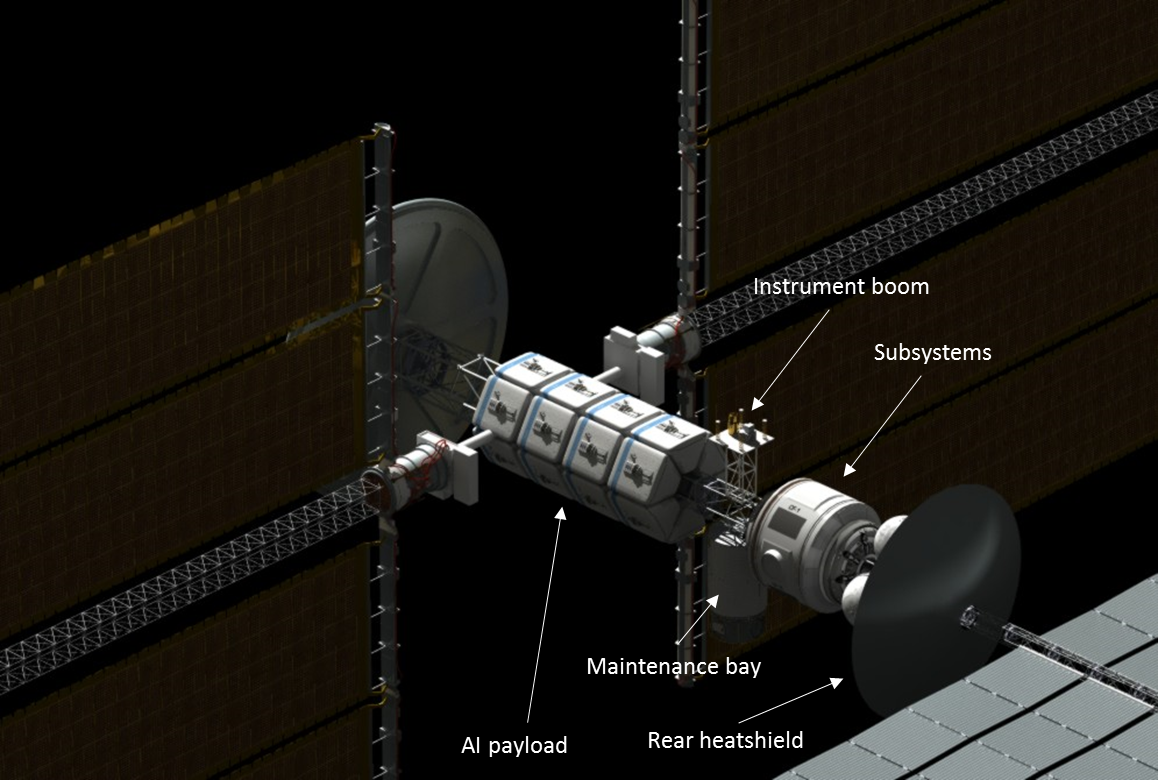}
\centering
\caption{View from the back of AI probe (Image: Adrian Mann)}
\label{fig:AIprobe2}
\end{figure}

Fig. \ref{fig:AIprobe1} and Fig. \ref{fig:AIprobe2} show an artist's impression of an AI probe with its main subsystems.

The design is similar to spacecraft with nuclear reactors but with important differences. Where the design of spacecraft with nuclear reactors is dominated by large radiators and placing the reactor far away from other spacecraft subsystems, the AI probe has solar cells and the computing payload does not need to be placed as far away from other subsystems for radiation concerns.

The AI payload is likely to have a cylindrical shape, as it is easier for the heat rejection system to have one backbone heat channel and then smaller, radial pipes that reject heat from the processing units. The heat is rejected via large radiators. The radiator size may decreases with distance from the payload, as less and less fluid is available for rejection (not shown in image). It is furthermore better to reject the heat quickly. Hence, the larger size of the radiators close to the payload. The radiator is perpendicular to the payload in order to avoid heat radiation from the payload being absorbed by the radiators and letting the payload face as much free space as possible.

Fig. \ref{fig:AIprobe2} shows an additional heat shield between the payload section and the radiators, in order to prevent radiative heat transfer from the payload. 
In order to maximize energy intake from the star, the spacecraft may be located as close as possible to the target star. There is likely to be a trade between distance to the star and other probe objectives. One can also imagine that sub-probes would do the majority of exploration and the AI probe would close to the star and do the majority of the computation-heavy tasks while communicating with the sub-probes. 

If the probe operates close to the target star, strong thermal radiation and particles from the star impact the spacecraft. In order to avoid heat and particle influx from the star, a heat and radiation shield is needed for protection against these particles. The shield is located in the direction of the star and shields the payload.

The spacecraft needs to be constantly maintained and parts replaced or repaired. This is similar to existing terrestrial supercomputers. A system of this complexity is very likely to need repair. If the computer is modular, these modules are replaced on a regular basis and parts replaced within these modules. We can imagine a storage depot of parts and robots that replace these parts. With more advanced technology available, robots that reproduce even very complex replacement parts can be imagined. 

The AI payload needs to either be protected against galactic cosmic rays during its interstellar cruise or needs to have appropriate counter-measures in place such as self-healing \cite{Han2016,Moon2016} and radiation-hardened electronics. 

\section{When will we be ready?}
\begin{figure}[h]
\includegraphics[height=5cm]{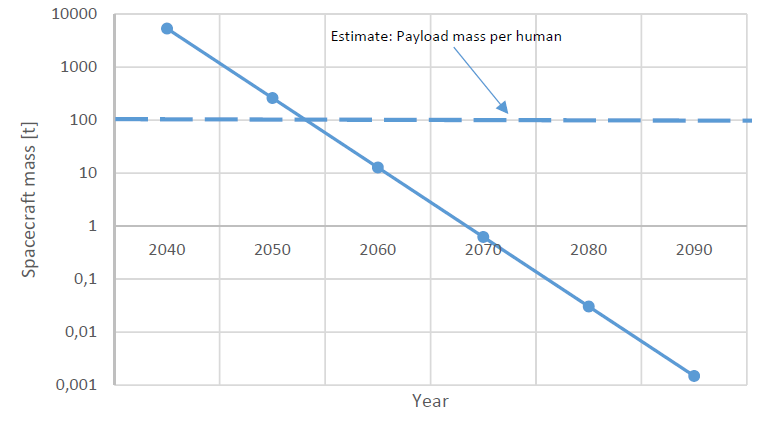}
\centering
\caption{Spacecraft payload mass vs. year of development}
\label{fig:AIpl}
\end{figure}

Under the assumption that during the 2050 to 2090 timeframe, computing power per mass is still increasing by a factor of 20.5, it can be seen in \ref{fig:AIpl} that the payload mass decreases to levels that can be transported by an interstellar spacecraft of the size of the Daedalus probe or smaller from 2050 onwards. If the trend continues till 2090, even modest payload sizes of about 1 kg can be imagined. Such a mission might be subject to the "waiting paradox", as the development of the payload might be postponed successively, as long as computing power increases and consequently launch cost decrease due to the lower payload mass. 

Furthermore, under the assumption that an advanced AI payload has equal capabilities for exploration than a human and the mass required for transporting a human over interstellar distances is estimated to be about 100t \cite{Hein2012b,Matloff2006}, the breakeven point for an AI probe with similar cognitive capabilities as a human would be somewhere between 2050 and 2060. However, it is clear that the similarities end here, as a human crew would have colonization as an objective and would also require a large number of crew members \cite{Smith2014b}.

\section{Conclusions}
We presented four types of artificial intelligence interstellar probes along with their required capabilities and mission architectures. Furthermore, a generic design for an artificial intelligence interstellar probe was presented. Based on the extrapolation of existing technologies and trends, we estimated that the payload of such an interstellar probe that has a similar computing power as the human brain is likely to have a mass of hundreds of tons in the 2050 time frame and a mass of dozens of tons in the 2060 time frame. Furthermore, estimates for the advent of artificial general intelligence and first interstellar missions coincide and are both estimated to be in the middle of the 21st century. We therefore conclude that a more in-depth exploration of the relationship between the two should be attempted, looking into currently neglected areas such as protecting the artificial intelligence payload from radiation in interstellar space and the role of artificial intelligence in self-replication. 

\bibliographystyle{alpha}
\bibliography{library.bib}

\end{document}